\documentclass[10pt]{article}

\usepackage{amsmath}
\usepackage{amssymb}

\usepackage{graphicx}

\usepackage{cite}

\usepackage{color} 

\usepackage[labelfont=bf,labelsep=period,justification=raggedright]{caption}

\makeatletter
\renewcommand{\@biblabel}[1]{\quad#1.}
\makeatother

\date{}

\begin{document}

% Title must be 150 characters or less
\begin{flushleft}
{\Large
\textbf{
What ecological factors shape species-area curves in neutral models?}
}
% Insert Author names, affiliations and corresponding author email.
\\
\bigskip Massimo Cencini$^{1}$, Simone Pigolotti$^{2}$, Miguel
A. Mu\~noz$^{3,\ast}$
\\
$^1$ Istituto dei Sistemi Complessi, Consiglio Nazionale delle
Ricerche, Via dei Taurini, 19 00185 Rome, Italy.\\
$^2$ Dept. de Fisica i Eng. Nuclear, Universitat Politecnica de
Catalunya Edif. GAIA, Rambla Sant Nebridi s/n, 08222 Terrassa,
Barcelona, Spain.\\
$^3$ Instituto Carlos I de F{\'\i}sica Te\'orica y Computacional,
Facultad de Ciencias. Universidad de Granada, 18071 Granada, Spain.\\
$^\ast$ E-mail:  mamunoz@onsager.ugr.es
\end{flushleft}

% Please keep the abstract between 250 and 300 words
\section*{Abstract}

Understanding factors that shape biodiversity and species coexistence
across scales is of utmost importance in ecology, both theoretically
and for conservation policies. Species-area relationships (SARs),
measuring how the number of observed species increases upon enlarging
the sampled area, constitute a convenient tool for quantifying the
spatial structure of biodiversity.  While general features of
species-area curves are quite universal across ecosystems, some
quantitative aspects can change significantly. Several attempts have
been made to link these variations to ecological forces.  Within the
framework of spatially explicit neutral models, here we scrutinize the
effect of varying the local population size (i.e. the number of
individuals per site) and the level of habitat saturation (allowing
for empty sites).  We conclude that species-area curves become
shallower when the local population size increases, while habitat
saturation, unless strongly violated, plays a marginal role. Our
findings provide a plausible explanation of why SARs for
microorganisms are flatter than those for larger organisms.

\section*{Introduction}
Species-area laws quantify the relation between area and the number of
species found in that area and represent one of the most robust
biodiversity patterns \cite{Drakare2006}. Clearly, larger areas
harbor a greater number of species, but the increase occurs in a
remarkably orderly way \cite{Schoener1976}.  Typically, empirical
species-area curves display an inverted $\mathcal{S}$-shape: at small
(local) and very large (continental) areas ($A$) the number of species
($S$) increases in a relatively steep (nearly linear) way with the
area, while the increase is shallower at intermediate areas
\cite{Preston1960,rosenzweig1995,Hubbell2001}. Whilst the two extreme
regimes are relatively easy to rationalize, the intermediate one
remains intriguing and has attracted much attention (see \cite{Drakare2006}
and references therein). Several fitting formulas have been
proposed to describe collected data \cite{rosenzweig1995,tjorve2003},
among which, the most widely adopted are the logarithmic law $S\sim
\ln(A)$ \cite{fisher1943} and the power-law relation
\cite{arrhenius1921}
\begin{equation}
S \propto A^{z}\,.
\label{eq:sar}
\end{equation}
Data from many field studies tend to slightly favor the power law fit
(\ref{eq:sar}) with values of the exponent $z$ showing a dependence on
environmental variables, e.g., the latitude
\cite{Drakare2006}. Moreover, body-size seems to be an important
factor in shaping SARs: with some provisos on possible biases due to
undersampling or taxa identification \cite{undersampling,Green2006},
species-area curves for microorganisms are typically shallower than
those of larger organisms \cite{horner2004,Green2004,Microbial}.
Different hypothesis have been put forward for the reduced spatial
diversification of microorganisms (see the review \cite{Green2006} and
references therein): enhanced dispersal rates due to large population
sizes and short generation times, decreased local diversification due
to low extinction rates (owing to large population sizes), and to low
speciation rates (because of horizontal gene transfer and imperfect
isolation). Despite the role of the local population size in
determining the mechanisms above, the effect of its variations has not
been tested (to the best of our knowledge) in the context of
individual based models.

Along with empirical studies, theoretical efforts have been devoted to
identify ecological mechanisms responsible for shaping species-area
curves \cite{rosenzweig1995}.  Examples of these mechanisms include
trade-off and interspecific competition \cite{Tilman1982,Chesson2000},
or predator-prey dynamics \cite{Brose2004} (see \cite{Lomolino} for a
review).  The \textit{Neutral theory} \cite{Hubbell2001} emphasizes
the role of {\it stochastic} mechanisms such as demographic processes,
able by themselves to generate nontrivial diversity patterns.  In
particular, neutral models incorporate processes such as colonization,
dispersal, and speciation and assume, in contrast with the
\textit{niche} paradigm \cite{chase2003}, that all individuals,
regardless of the species they belong to, have the same prospects of
death, reproduction, etc.

{\it Spatially implicit} neutral models have been shown to produce
species abundance distributions (SADs) in remarkable good agreement
with empirical data \cite{Hubbell2001,Volkov2003}. This suggests that
they capture the essence of general and robust community-level
properties or, at least, promotes neutral theories to suitable
null-models \cite{bell2001}.  Etienne et al.\cite{Etienne2007} showed
that SADs remain unaltered when breaking Hubbell's \cite{Hubbell2001}
``zero-sum assumption'', postulating that the community size is
strictly kept constant by resource saturation. Then, the question
arises of whether the spatial distribution of species is equally
robust upon modifying other ``details'' of the underlying neutral
theory?  (see \cite{Harte2009}). If not, what are the relevant
ecological mechanisms/forces that, implemented in a neutral model, are
relevant for shaping the SARs and thus the value of $z$ as.  For
example, what is the relevance of body-size?

{\it Spatially explicit} neutral models generate species-area curves
very similar qualitatively and, to some extent, quantitatively, to
empirical ones. They display power-law behaviors with an exponent $z$
in a realistic range \cite{Durrett-Levin,Chave2002}.  Species-area
curves in spatially explicit neutral models are mainly shaped by the
interplay of dispersal limitation and speciation
\cite{Bramson1996,Durrett-Levin}.  In particular, for finite ranged
dispersal kernels, regardless their specific form, the actual value of
$z$ is mainly determined by the speciation rate
\cite{Rosindell2007,Pigolotti2009}, which is however difficult (or
impossible) to estimate.  Sensitive variations of the exponent value,
at fixed speciation rate, have been observed when the dispersal
process couples distant locations in the ecosystem, e.g. by
considering fat tailed distributions \cite{Rosindell2009}.  The
influence of other factors was investigated by Chave et
al. \cite{Chave2002} who mainly focused on violations of the neutral
assumptions, e.g., by introducing trade-offs.

In this paper, we study the effect of varying the number of
individuals that can live at a single ecosystem site on the
species-area curves generated by neutral spatial models.  We consider
two kinds of variations: allowing for large \textit{local population
  sizes}, by letting each site host many individuals, as appropriate
for describing communities of microorganisms connected by dispersal,
and allowing for empty sites, i.e.  changing the level of
\textit{habitat saturation}.

To explore these possibilities, we present extensive simulations of
the stepping stone model (SSM) \cite{Kimura1953,DurrettBook}, which
incorporates variable local population size by increasing the number
of allowed individuals per site, and the multispecies (or multitype)
contact process (MCP) \cite{Liggett1985}, which is suited to study non
saturated habitats.  These models have been not thoroughly explored
before in the context of spatial neutral theory: in particular, the
MCP, discussed by Durrett and Levin \cite{Durrett-Levin}, was not, to
the best of our knowledge, previously simulated. The SSM is popular in
the context of population genetics but its predictions for species
area laws have not been explored before.  To complete the picture, we
compare the species-area relationships generated by the above models
with those for the multispecies voter model (MVM), which is possibly
the most studied spatially explicit neutral model
\cite{Durrett-Levin,Rosindell2007,Pigolotti2009}.  We remark that the
term ``voter model'' is often used to denote the model with
nearest-neighbor dispersal among sites. In this paper, we use the same
name also when more general dispersal kernels are considered.

Common to all the above models is that individuals of different
species are placed at the sites of a two-dimensional lattice and
evolve according to basic demographic processes such as birth, death,
migration, and speciation. However, important differences also exist.
While the MVM and SSM describe saturated habitats with a constant
density of individuals, the MCP describes fragmented systems where the
density of individuals is irregular both in space and time, with the
presence of gaps. The models differ also in the number of allowed
individuals per site. In the MVM and MCP each site can hosts one
individual at most, as appropriate to describe large organisms, such
as trees.  In the SSM, each site represents a local community of $M$
individuals, making the model more suitable to describe, e.g., patches
of microorganisms connected by migration \cite{Vanpeteghem2010}.
Indeed, as discussed by Fenchel and Finlay \cite{Fenchel2004-2},
comparing larger and smaller organisms is essentially equivalent to
compare organisms with smaller and with larger population sizes,
respectively.  For instance, it has been estimated that one gram of
typical soil can host $10^6-10^7$ bacteria \cite{Whitman1998}.

We conclude that --together with speciation rate and the dispersion
kernel-- the size of the local population is an important shaping
factor for neutral predictions on species spatial distributions and,
hence, on SAR curves. On the other hand, mild violations of habitat
saturation --- i.e. not as extreme as to break the space into isolated
regions --- have little effect on the slope of SAR curves on scales
larger than the typical size of the gaps.

\section*{Methods}
We now present the three aforementioned spatially-explicit neutral
models and discuss afterward their main similarities and differences.
The section is organized as follows. In the three
  first subsections we introduce and motivate the models that will be
  the subject of our study. Then we discuss their similarities,
  differences and numerical implementation. The last subsection is
  devoted to a discussion of the effect of the choice of the dispersal
  kernel.

\subsubsection*{Multispecies voter model (MVM)}
The multispecies voter model is a spatial generalization of the
infinite allele Moran model used in population genetics (see,
e.g. \cite{Gillespie}).  Each site of a square lattice is always
occupied by a single individual: the habitat is thus saturated.  At
each time step, a randomly chosen individual on the lattice is killed
and immediately replaced: with probability $(1-\nu)$,
by a randomly chosen copy of one of the nearest
  neighbors (dispersal event); with probability $\nu$, by an
individual from a new species (speciation event).  When $\nu>0$, any
species will eventually go extinct; speciation events compensate
extinctions so that a dynamical equilibrium eventually sets in
\cite{Durrett-Levin}.

\subsubsection*{Stepping stone model (SSM)}
In the previous model, each lattice site hosts a single individual, as
appropriate when modeling, e.g. a forest, where each site represents
the space occupied by a single tree. In such cases, the limiting
resources are indeed strongly related to space, so that it is
reasonable to model competition by simply assuming that when an
individual dies, a vacant site is left to be occupied by another
individual. Conversely, microorganisms, such as small eukaryotes or
bacteria, are often present in very large numbers below a scale in
which one can assume that all individuals share the same pool of
resources.  Therefore, it is more appropriate to think of the habitat
as subdivided into small patches, connected by migration and each
hosting a large population of individuals that directly compete with
each other \cite{Fenchel2004-2,Vanpeteghem2010}.  Such a setting is
even more relevant when the habitat is physically divided into
patches, so that moving from a patch to another is more difficult than
moving within a patch, like in the case of an island chain or of soil
fragmented in different soil grains.  \cite{warren}.

  In this perspective, the stepping stone model, originally introduced
  in population genetics \cite{Kimura1953}, straightforwardly
  generalizes the MVM by allowing each site to host a fixed (but
  arbitrary) number, $M$, of individuals.  At each time step an
  individual is randomly selected, killed and then replaced: with
  probability $(1-\nu)$ by the offspring of an existing individual or,
  with probability $\nu$ by an individual of a new species. In the
  former case, the parent is chosen with probability $(1-\mu)$ among
  the remaining $M-1$ individuals residing at the same site and with
  probability $\mu$ among those at a randomly chosen nearest neighbor
  site.  For $M=1$, the SSM recovers the MVM with (as detailed in
  Appendix S2) $\nu$ substituted by $\bar{\nu}=\nu/\mu$, which is the
  effective speciation-to-diffusion ratio in the SSM.  Note that, in
  general, as dispersal occurs every $1/\mu$ time steps and not at
  every time step as in the other models, one can show that comparison
  should be done equating $\bar{\nu}=\nu/\mu$ (speciation-to-migration
  rate) to the value of $\nu$ used in the other models (see, e.g,
  \cite{DurrettBook}).

\subsubsection*{Multispecies contact process (MCP)}
In the MVM, gaps left by deaths are immediately filled by newborns
leading to habitat saturation. This is tantamount to assuming
reproduction rates infinitely larger than death rates
\cite{Durrett-Levin}.  In the contact process \cite{Liggett1985}, this
assumption is relaxed and gaps can survive for arbitrarily large
times. In particular, each individual dies at rate $\delta$ and
reproduces at rate $\beta$ giving rise to a newborn at a randomly
chosen neighbor site. As each site cannot host more than one
individual, attempted reproduction is successful only if an empty
neighbor is chosen.  When reproduction is successful, the newborn
belongs to the parent species with probability $(1-\nu)$ and to a new
species with probability $\nu$.  Thus, the relevant parameters are the
speciation rate $\nu$ and the birth-to-death ratio
$\gamma=\beta/\delta$, controlling the fraction of occupied sites in
steady state conditions.  For large $\gamma$, gaps are small and
infrequent: in the limit $\gamma \rightarrow \infty$ the MCP recovers
the MVM \cite{Durrett-Levin}.  Conversely, lowering $\gamma$ results
in an unsaturated habitat with larger and longer-lived gaps. Finally,
at $\gamma\lesssim 1.649$, i.e. the CP critical point
\cite{MarroBook}, births become too infrequent, leading to global
extinction.

\subsection*{Similarities and differences between models}

We now discuss the main similarities and differences among the above
models, as summarized in Table~\ref{tab:models}.  A key feature is the
maximum number of individuals allowed at each site: $1$ for the MVM
and the MCP, $M$ for the SSM.  While the MVM and the SSM describe a
saturated habitat, in the MCP, as sites can be empty, the habitat is
not saturated.  In all models, diversification is implemented as point
speciation \cite{Hubbell2001}, which of course should not be regarded
as a realistic speciation mechanism but rather as an effective one
\cite{Kopp2010}. In this perspective, the speciation rate $\nu$ has to
be interpreted as a normalized rate (speciation over death rate).
Moreover, as said above, due to the different dispersal rule, in SSM
the proper quantity to set up a comparison with the other two models
is the speciation to migration ratio $\bar{\nu}=\nu/\mu$.
\begin{table}[!ht]
\caption{\textbf{Summary of models}}
\begin{tabular}{|l|c|c|} 
\hline 
model & local population & saturation \\ %%& interactions 
\hline 
MVM & $1$ & Y  \\ 
\hline
MCP &  $\{0,1\}$ & N  \\ 
\hline
SSM & $M$ & Y  \\ 
\hline
\end{tabular}
\begin{flushleft}
Summary of the main features of the considered
  spatially-explicit neutral models. Y/N stands for
  Yes/No.
\end{flushleft}
\label{tab:models}
\end{table}

Concerning the simulation scheme, the voter model
  and the stepping stone model can be reformulated in terms of
  coalescent random walkers \cite{Liggett1975}, leading to approximate
  estimates of the exponent $z$ (that, for MVM, were put forward in
  \cite{Bramson1996,Durrett-Levin}) and also to very efficient
  numerical implementations
  \cite{Rosindell2007,Rosindell2008,Pigolotti2009}.  One of the main
  advantage of this method is that numerical simulations are virtually
  free from boundary effect problems as if simulating a portion of an
  infinite landscape \cite{Rosindell2007}. Details on the coalescing
  random walk analogy, the resulting numerical scheme and analytical
  estimates are discussed in Appendices S1, S2 and S3. Unfortunately,
  such reformulation does not easily extend to the multitype contact
  process, which was simulated by means of a standard algorithm
  \cite{MarroBook} adapted to the multitype case.  In this case,
  periodic boundary conditions have been employed and tests to
  minimize possible finite size effects performed. Appendix S4 details
  the numerical scheme.

To close this section, we remind that, while in this
  paper we restricted our comparison to models in which competitive
  interactions among individuals are present, recently, O'Dwyer and
  Green \cite{OG2010} introduced a model in which individuals do not
  compete (so that species are independent).  In this case the number
  of individuals per site is unrestricted; the advantage of this
  simplifying assumption is that it allows for a full analytical
  treatment of the problem.

\subsection*{Dispersal kernel and species-area relationships}

The above models have an additional degree of freedom
  related to the choice of the dispersal kernel, which is, in general,
  important to reproduce SAR curves similar to empirical ones.  In
  particular, nearest neighbor (NN) kernels generate biphasic SAR
  curves rather than triphasic ones \cite{Chave2002}, because the
  steep-growth regime at small areas cannot be reproduced.  To observe
  triphasic $\mathcal{S}$-shaped SAR, similar to empirical ones,
  requires more general (finite-range) dispersal kernels, acting on
  several sites. Moreover, the resulting SAR curves do not depend on
  the shape of the kernel but only on its range \cite{Rosindell2007}.
  Fat-tailed dispersal kernels could also be considered to model some
  dispersal mechanisms found in nature, and have been found to
  quantitatively influence SARs both in terms of the extension of the
  intermediate range and in terms of the exponent $z$ values
  \cite{Rosindell2009}.

In this paper we mostly explored the behavior of the
  species-area curves by implementing the above described models with
  the nearest-neighbor kernel. This choice is mainly dictated by its
  simplicity and by the costs of simulating the SSM with large
  lattices (as necessary if more general kernels are used) when the
  local population size becomes large. Moreover, for the MVM, at small
  areas SAR curves obtained with NN-kernel approximatively behave as
  power laws and qualitatively match the behavior of the intermediate
  (power law) regime of more general kernels \cite{Pigolotti2009}.

However, to test the robustness of our main findings
  against the kernel choice we also performed simulations by employing
  a finite-range square kernel: a killed individual at a given site
  can be replaced by any of the individuals present in a square
  centered at that site and having side $2K+1$. We remind that for the
  MVM, as soon as $K\gtrsim 5$, $z$ does not depend on $K$
  and the entire curve can be rescaled
  \cite{Rosindell2007,Pigolotti2009}. For this reason tests have been
  performed at $K=7$.

% Results and Discussion can be combined.
\section*{Results}

The speciation rate $\nu$ determines most features of neutral
species-area curves, in particular, the interesting power-law regime
Eq.~(\ref{eq:sar})
\cite{Durrett-Levin,Chave2002,Zillio2005,Rosindell2007,Pigolotti2009}.
Therefore, to discriminate the influence of the different ecological
mechanisms incorporated in the models, we will compare species-area
curves obtained by the above introduced models at equal $\nu$.

Once $\nu$ (and the dispersal kernel) are fixed, the MVM is fully
specified, while the SSM and the MCP need additional parameters to be
set.  As previously shown, both the SSM and the MCP reduce to the MVM
for $M=1$ and large birth over death rate ratio
($\gamma\!=\!\beta/\delta\gg 1$), respectively. Hence, to stay away
from this limit, we allowed for a large local community size for the
SSM by choosing $M=100$ with migration probability $\mu\!=\!0.1$
(holding $\bar{\nu}\!=\!\nu/\mu$ equal to $\nu$ in the other models,
as specified above), and considered habitat unsaturated conditions for
the MCP by choosing $\gamma=1.68$, ensuring that only $\approx
0.095\%$ of the available sites are occupied.

\begin{figure}[!ht]
\begin{center}
\includegraphics[width=4in]{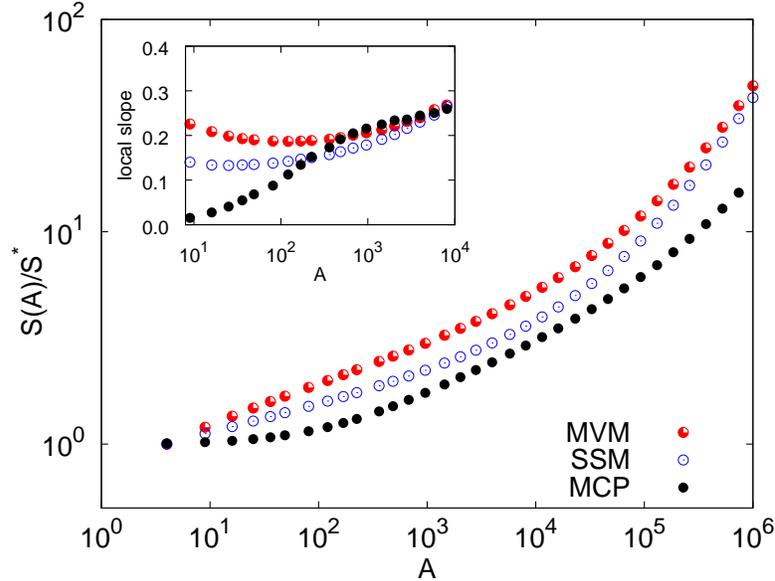}
\end{center}
\caption{ \textbf{ SARs generated by the three models at
    $\nu\!=\!10^{-6}$ ($\nu/\mu\!=\!10^{-6}$ for the SSM) using
    nearest neighbor dispersal.}{The MVM and the SSM are simulated on a
    $1000\times 1000$ square lattice. In the SSM, we chose $M=100$ and
    $\mu=0.1$. The MCP is simulated on a $2500 \times 2500$ lattice
    with $\gamma=1.68$ adopting weighted averages (see
    Fig.~\ref{fig:CP} and related discussion).  In all cases, we
    averaged over $5\times 10^2-10^3$ independent realizations and
    statistical errors are smaller than symbols' size.  To ease the
    comparison, $S$ has been normalized by the average number of
    species $S^*$ at the smallest sampled area.  Inset: local slopes
    $d\ln S/d\ln A$ of the four curves for areas smaller than
    $10^4$. } }
\label{fig:SARcompare}
\end{figure}

To compare species-area curves generated by the three models, we
performed extensive numerical simulations of the MCP and the SSM (see
Appendices S2 and S4 for the numerical implementation).
Most of the simulations have been performed by using
  nearest-neighbor kernels and tests on SSM and MVM have been done
  using the square kernel discussed in \textit{Methods}. For the MVM
we relied on already published numerical results \cite{Pigolotti2009}.
Figure~\ref{fig:SARcompare} shows the species-area curves generated by
the three models at $\nu$ ($\bar{\nu}$ in the SSM) equal to
$10^{-6}$. The curves are qualitatively similar to each other.  They
display a shallower than linear growth for small areas and become
steeper, eventually linear, at larger areas. The transition between
these two regimes occurs at a similar scale (shown to be $O(1/\nu)$
for the MVM \cite{Durrett-Levin}) in all models.

The interesting regime can be quantitatively scrutinized by looking at
the local slopes, $\mathrm{d}\ln S/\mathrm{d}\ln A$, shown in the
inset of Fig.~\ref{fig:SARcompare} for small areas.  A power-law
range, as in Eq.~(\ref{eq:sar}), would correspond to a region in which
$\mathrm{d}\ln S/\mathrm{d}\ln A \approx const=z$.  This is a good
approximation for the MVM and the SSM whose local slopes are
characterized by a shallow parabolic shape. As customary in recent
literature \cite{Rosindell2007,Pigolotti2009,OG2010}, in the
following we shall determine the exponent $z$ as the minimum of this
parabola; equivalent (within error bars) results can be obtained
fitting a power-law as Eq.~(\ref{eq:sar}) on the species-area curve in
the scaling range.

Local slopes and thus $z$ display some variability among the three
models. In particular, the stepping stone model gives rise to
shallower curves with respect to the voter model,
i.e. $z_{_\mathrm{SSM}}<z_{_\mathrm{MVM}}$. On the other hand, no
clear power-law range can be identified for the MCP, as the local
slope increases monotonically from zero at increasing the area.  We
anticipate that this behavior is due to the presence of gaps in the
distribution of individuals (see the subsection
  \textit{Multispecies Contact Process} below).

\begin{figure}[!ht]
\begin{center}
\includegraphics[width=4in]{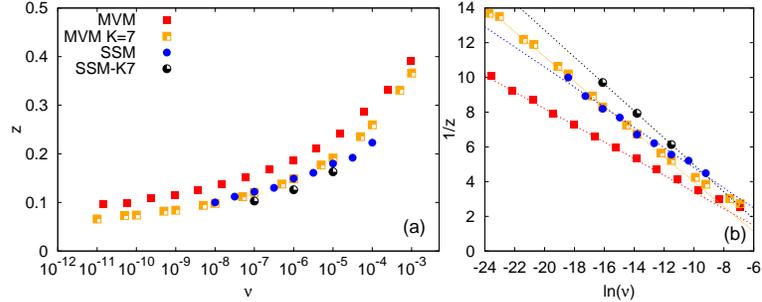}
\end{center}
\caption{ \textbf{Exponent $z$ as a function of $\nu$
      for the MVM and SSM. } Panel (a) shows $z$ vs $\nu$ for MVM and
    SSM with NN- and square-kernel with $K=7$. The SSM is simulated
    for $M=100$ and $\mu=0.1$.  Due to computational limitation,
    simulations for the SSM at $K=7$ have been performed at three
    different values of $\nu$ only. The system size has been chosen in
    each simulation in order to properly resolve the power-law regime.
    Panel (b) shows $1/z$ vs $\ln(\nu)$ for the same data of
    (a). Dotted straight lines are best fits obtained using
    Eq.~(\ref{eq:fit}). Fitted values are: for the NN-kernel MVM
    $m\approx -0.48$ $q\approx -1.4$ and SSM $m\approx -0.58$
    $q\approx -1$; for the square-kernel $K=7$ MVM $m\approx -0.72$
    $q\approx -3.2$ and SSM $m\approx -0.78$ $q\approx -2.8$}
\label{fig:z-vs-nu}
\end{figure}

Figure~\ref{fig:z-vs-nu}a shows the dependence of the
  exponent $z$ on the speciation rate $\nu$ ($\tilde{\nu}$ for SSM)
  for the MVM and SSM; MCP was excluded because as seen in
  Fig.~\ref{fig:SARcompare} no reasonable power-law range exists for
  $\gamma$ close to $\gamma_c$. Let us start comparing the two models
  with NN dispersal.  As for the case $\nu=10^{-6}$
(Fig.~\ref{fig:SARcompare}), the exponents are different and the
curves produced by the SSM are consistently shallower than those
generated by the voter model in the explored range of $\nu$-values.
In this figure we can see that the exponents for the
  SSM with NN-dispersal appear to be close to (but not coincident
  with) those of the MVM with the square-kernel ($K=7$). However, when
  comparing the exponents of the SSM and MVM when the square-kernel
  ($K=7$) is employed for both, we still observe that the former is
  shallower (see also Fig~\ref{figure-SSM} and its discussion in the
  next section).  Notice that increasing further $K$ in the MVM does
  not produce further changes in the exponent
  \cite{Rosindell2007,Pigolotti2009}.  Therefore, as the comparison
  with the same dispersal kernel reveals, the decrease in the exponent
  $z$ due to the increase of the local population size is a genuine
  effect. We also observe that the function $z(\nu)$ is remarkably
similar in the two models (independently of the
  dispersal kernel employed), as demonstrated in
Fig.~\ref{fig:z-vs-nu}b where $1/z$ is shown as a function of $\ln
\nu$.  In particular, both models are fairly well described by the
fitting formula \cite{Pigolotti2009}
\begin{equation}
1/z = q + m \ln(\nu) \,,
\label{eq:fit}
\end{equation}
where the constants $q$ and $m$ are model-dependent.

For the MVM (with NN-dispersal), some mathematical
results are available for $z(\nu)$.  Durrett and Levin
\cite{Durrett-Levin} (see also \cite{Bramson1996}) provided the
asymptotic estimate $z_{_\mathrm{MVM}}(\nu)\approx {(2
  \ln(\ln(\nu^{-1}))-\log(2\pi))}/{ \ln(\nu^{-1})}$, which is
consistent with the fitting formula (\ref{eq:fit}), but for a very
slow variation of the slope $m$ due to the $\ln\ln(\nu^{-1})$
term. The specific values of $m$ and $q$ obtained by the numerical
simulations are slightly different from those implied by the
asymptotic estimate (see Ref.~cite{Pigolotti2009} for a detailed
discussion). For the SSM, as described in Appendix S3, we derived the
approximate asymptotic formula
\begin{equation}
z_{\mathrm{SSM}}(\nu) \approx \left\{
\begin{array}{ll}
z_{_\mathrm{MVM}}(\nu) & M\mu\ll 1 \\
z_{_\mathrm{MVM}}(\nu)/2 & M\mu\gg 1 
\end{array}
\right.\,.
\label{eq:analytics}
\end{equation}
Consistently with our numerical results, the above estimate predicts
that in the limit of large local population sizes ($M\mu\gg 1$) the
species-area curves of the SSM are shallower than those of the voter
model, which are recovered in the limit of small local population
sizes ($M\mu\ll 1$).  We also mention that the fitting formula
Eq.~(\ref{eq:fit}) is also compatible with the result of an exactly
solvable variant of the neutral model \cite{OG2010}.

The following two sections focus on the SSM and the MCP, to further
elucidate the importance of local community size and habitat
saturation on the variability of SAR curves.

\subsection*{Multispecies Stepping Stone Model}

\begin{figure}[!ht]
\begin{center}
\includegraphics[width=3in]{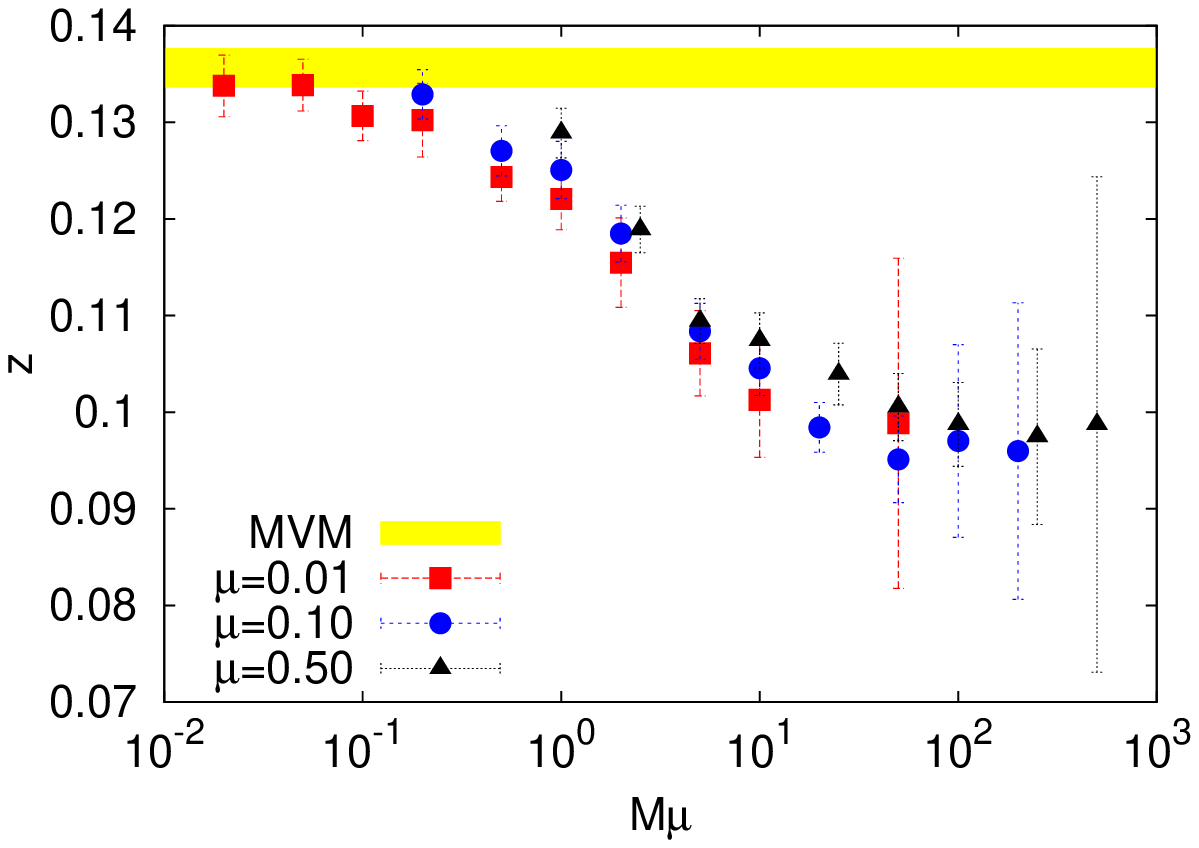}
\includegraphics[width=3in]{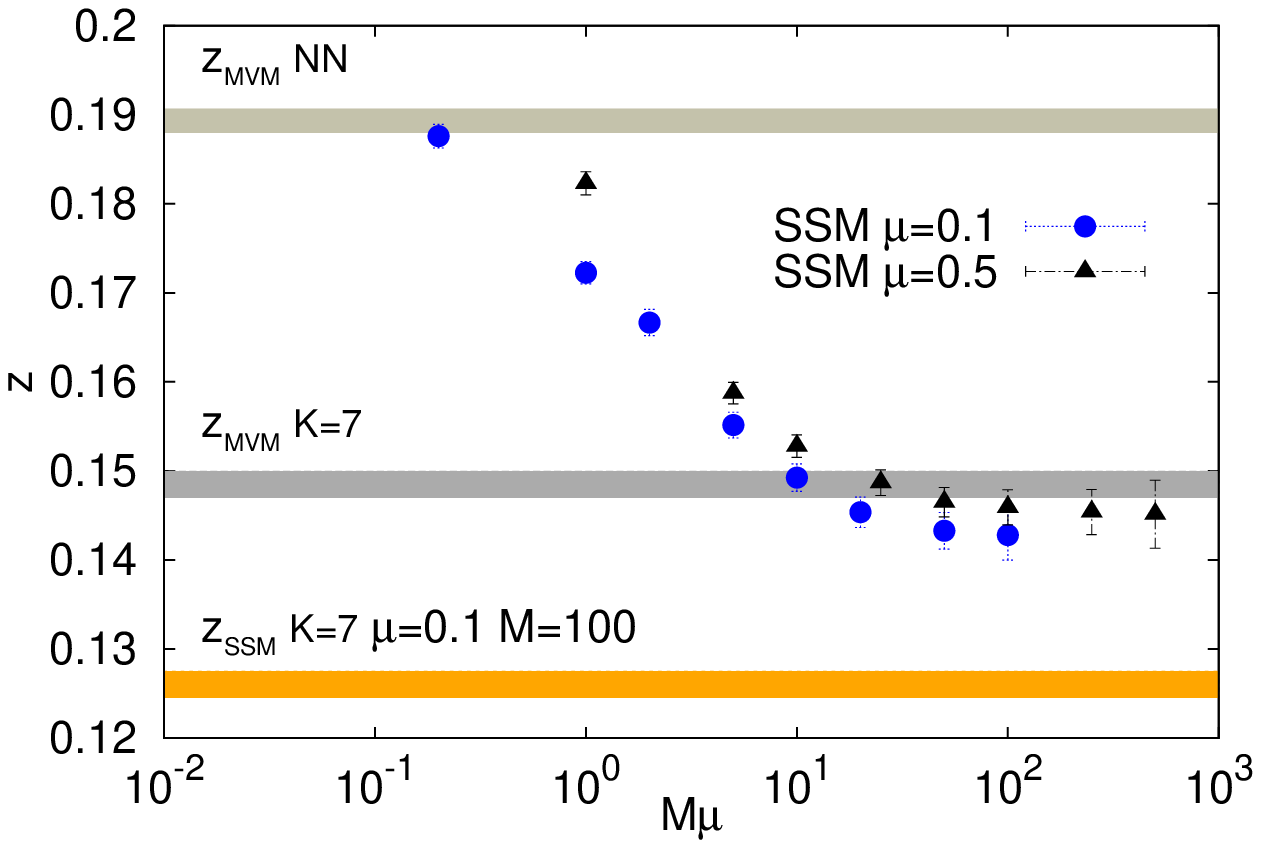}
\end{center}
\caption{ \textbf{ Exponent $z$ for the SSM as a
      function of $M\mu$ at fixed $\bar{\nu}=\nu/\mu$.}  The left
    panel shows the case $\bar{\nu}=\nu/\mu=10^{-8}$ for
    $\mu=0.5,0.1,0.01$ with NN-kernel. The shaded area indicates the
    value $z_{_\mathrm{MVM}}$ (including the estimated error) of the
    exponent for the MVM with $\nu=10^{-8}$.  For $M\mu\gg 1$
    statistical errors increase because a smaller number of
    realizations was used as simulations become very costly. The right
    panel shows the case $\bar{\nu}=\nu/\mu=10^{-6}$ for $\mu=0.5,0.1$
    with NN-kernel. The shaded areas display the value of the exponent
    for the MVM with both the NN- and square-kernel ($K=7$) and the
    SSM with square-kernel ($K=7$) for $M=100$ and $\mu=0.1$
    (i.e. $M\mu=0.1$).}
\label{figure-SSM}
\end{figure}

Sensitive variations of $z$ are indeed observed by changing $M$ and
$\mu$, for fixed speciation to migration ratio $\bar{\nu}=\nu/\mu$. In
particular, the exponent $z$ decreases with $\mu$ and $M$ and seems to
be mainly determined by their product $ M \mu$, as shown in
Fig.~\ref{figure-SSM} for two different values of $\bar{\nu}$. 

For $M\mu\ll 1$ the exponent $z$ approaches the corresponding value in
the MVM, while at large $M\mu$ the exponent decreases in a sigmoidal
fashion and displays a tendency towards a different asymptotic value.
These two limits correspond to very different regimes. When $\mu M$ is
very small, sites have a small local population and species
(individuals) exchanges among sites are rare: most sites are not able
to sustain diversity and contain only one species (i.e. in
  this regime local fixation dominates). In this limit, the SSM
reproduces MVM behavior with the on-site mono-dominant community in
the former playing the role of a single individual in the
latter. Conversely, when $\mu M$ is very large and $\nu$ is very
small, the large local community size (buffering
  local extinctions and fixations) and the frequent exchanges among
sites allow each site to host a large number of species on average. A
further consequence is that each species will be statistically
represented in a similar way at each site of a large region. Also,
distant sites can now host many common species. This leads to
shallower species-area curves, and thus to the smaller $z$ values
shown in Fig.~\ref{figure-SSM}. It is worth remarking that shallower
SARs do not necessarily mean lower diversity as, for $M\mu \gg 1$, the
prefactor in front of the power-law (\ref{eq:sar}) can be very large
(not shown here).

Remarkably, the above qualitative argument can be supported by
analytical estimates, see Eq.~(\ref{eq:analytics}). By generalizing
the calculation of Durrett and Levin \cite{Durrett-Levin} (see also
\cite{Bramson1996}), we have been able to estimate that, for $M \mu
\gg 1$, one should observe $z_{_\mathrm{SSM}}\approx
z_{_\mathrm{MVM}}/2$ (see Appendix S3 for details). The numerical
results of Fig.~\ref{figure-SSM} display the correct tendency: for the
largest values of $M\mu$ we could explore, we observe that $z$ is
reduced by a factor $\approx 1.4$ with respect to $z_{_\mathrm{MVM}}$.
This behavior is also confirmed for varying values of $\bar{\nu}$ (not
shown).  It would be very interesting to explore the (numerically
costly) larger values of $M\mu$ to test the theoretical prediction.

We close this section observing that in the right
  panel of Fig.~\ref{figure-SSM} we also show the value of the
  exponent $z$ obtained by using the square kernel for both the MVM
  and the SSM (we only show the value for $M\mu=10$ here as it is
  already close to the saturation regime). As already mentioned while
  the MVM with the square kernel is not far from the values of the
  exponent obtained with the SSM with NN-dispersal, still the exponent
  for the SSM with square kernel is sensitively smaller, confirming
  the robustness of the effect of increasing the local population
  size. 

\subsection*{Multispecies Contact Process}

\begin{figure}[!ht]
\begin{center}
\includegraphics[width=4in]{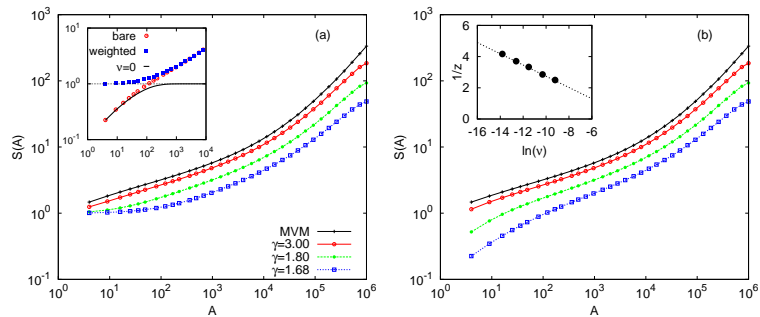}
\end{center}
\caption{
\textbf{SARs for the MCP for various
  $\gamma$ as labeled: (a) with weighted averages Eq.~(\ref{weighted})
  and (b) with bare averages Eq.~(\ref{blind})}.  Inset of (a): for
  $\gamma=1.68$, comparison among bare and weighted averages for
  $\nu=10^{-5}$ with the bare average for $\nu=0$. Notice the
  flattening of the weighted curve at short scales and the equivalence
  of both averages at larger scales. Inset of (b): $1/z$ vs
  $\ln(\nu)$, the straight line shows formula Eq.~(\ref{eq:fit}) with
  fitted values $m=-0.36$ and $q=-0.86$.  The exponents $z$ were
  estimated as the minimum of the local slopes
  of the species-area curves.
}
\label{fig:CP}
\end{figure}

At fixed speciation rate and varying the birth-to-death ratio
$\gamma\!=\!\beta/\delta$ of the MCP, we can inspect how the level of
habitat saturation affects SAR-curves. For $\gamma \!\gg \!\gamma_c
\!\approx\! 1.649$, the habitat is close to saturation, as the density
of occupied sites approaches $1$, and the MCP is equivalent to
MVM. Indeed, as shown in Fig.~\ref{fig:CP}, the curves generated by
the two models are essentially coincident already for $\gamma\!=\!3$.

For highly non-saturated habitats, i.e. for smaller values of
$\gamma$, larger and larger areas with very few (or zero) individuals
become more and more probable. In this regime, SAR curves display a
strong dependence on the choice of the sampling procedure, as we
illustrate here with two examples.

The first procedure, that we dub ``bare average'', consists in
ignoring the non-saturation of the habitat and thus averaging over
many samples of fixed area $A$, regardless of the number of
individuals they host:
\begin{equation}
  S(A)= \frac{\sum_{i}^{N(A)} s_i }{N(A)} ,
  \label{blind}
\end{equation} 
where $s_i$ is the number of distinct species in sample $i$ and $N(A)$
is the total number of samples of area $A$.  With this procedure,
$S(A)$ will be inevitably affected by the spatial variations of the
density of individuals. For instance, for $A=1$ (i.e. on a single
site), $s_i=1$ or $s_i=0$ so that $S(1)$, as given by
Eq.~(\ref{blind}), reduces to the average density.

A second and more appropriate choice is to put less weight on areas
with a smaller number of individuals, where the number of observed
species is statistically biased to be small.  In particular, by
denoting with $n_i$ the number individuals present in the area $a_i$,
we define the ``weighted average'' (which was used in
Fig.~\ref{fig:SARcompare}) as
  \begin{equation}
  S(A)= \frac{\sum_{i}^{N(A)} s_i n_i}{\sum_i^{N(A)} n_i}\,.
  \label{weighted}
\end{equation}

SAR curves for different values of $\gamma$ are shown in
Fig.~\ref{fig:CP}a and \ref{fig:CP}b for weighted and unweighted
averages, respectively.  For large $\gamma$, the dependence on the
averaging protocol (if any) is mild, while it is strong for small
values of $\gamma$. The strongest effects are observed at small areas,
where bare averages Eq.~(\ref{blind}) are influenced the most by local
densities.  This effect is demonstrated in the inset of
Fig~\ref{fig:CP}a, where the black line is simply the fraction $P(A)$
of regions of area $A$ occupied by at least one individual (of any
species). For small areas, the weighted average becomes very shallow
without any signature of power-law behavior (see inset of
Fig.~\ref{fig:SARcompare}). Conversely, the bare average almost
coincides with $P(A)$, demonstrating its lack of sensitivity to the
presence of more than one species at small scales. At larger scales,
where $P(A)\sim 1$, the two averages coincide.

In contrast with weighted averages and similarly to the other models
in Fig.~\ref{fig:SARcompare}, the local slopes of bare averages
display a parabolic intermediate range with a well defined minimum,
from which we can extract an estimate of exponent $z$.
Fig~\ref{fig:CP}b shows $1/z$ as a function of $\ln(\nu)$ for
$\gamma=1.68$; formula (\ref{eq:fit}) fits well these data, yielding
values of $z$ being larger that those for the MVM and the SSM. However,
as detailed in Appendix S5, in this case the interpretations of these
exponents is problematic as the power-law can be induced by the
spatial fluctuations of the density of individuals rather than by
species distribution.

\section*{Discussion}
In this paper, we have studied the effect of changing the level of
habitat saturation and the local population size on spatial neutral
models. We have shown that species-area laws quantitatively depend on
these ecological features, which go beyond the previously explored
variations due to long-ranged dispersal kernels \cite{Rosindell2009}.
Spatially explicit neutral models thus seem to be much richer in
structure than spatially implicit ones, where the species-abundance
distribution seems to be insensitive to implementation details.
Moreover, the observed variations of SARs suggest that spatial neutral
theories can explain part of the variability of the exponent $z$
observed in nature.

In spatially explicit neutral models, SAR curves typically display a
range of scales where they are well approximated by the power law
(\ref{eq:sar}), in particular at small scales for NN-kernels and at
intermediate scales for finite range kernels.  We have shown that,
generally, the inverse exponent $1/z$ is very well described as a
linear function of $\ln(\nu)$, the logarithm of the rate of the
introduction of new species $\nu$. The same kind of behavior was
analytically confirmed in an exactly solvable neutral model
\cite{OG2010}. However, the coefficients of this linear relation and
thus the actual value of $z$ are sensitive to the ecological factors
implemented in the model.  The logarithmic behavior is a general and
robust feature related to two common features of all neutral models
discussed here: species originate with one individual (point
speciation mode) and then diffuse in space. Altering the speciation
mechanisms in spatially implicit models affects some aspects of SADs
\cite{Etienne2007m,Rosindell2010,Etienne2011}.  It would be
interesting, in the future, to study how different speciation modes
reflect into the spatial variation of biodiversity, a program which
just started in the context of spatial models (see \cite{Kopp2010} and
references therein). In the context of the models considered in this
paper, assuming that variations in $z$ are caused by $\nu$ variability
among different taxa, amounts to say that the diversification rate per
capita per generation increases at increasing body mass
\cite{Pigolotti2009}. While this possibility cannot be completely
ruled out (owing mostly to the difficulty of estimating such rates),
organisms such as bacteria are characterized by high mutation rates
and genetic plasticity, rather suggesting a higher rate of
differentiation even when considered at the individual level.

Relaxing the hypothesis of \textit{habitat saturation} --as occurs in
the multispecies contact process-- does not greatly modify the
behavior of species-areas curves with respect to the saturated case
--i.e. the multispecies voter model--, unless the habitat becomes too
fragmented.  In the latter case, species-area curves strongly depend
on the sampling procedure.  In particular, using ``weighted averages''
(which weight to the sampled area proportionally to the population it
hosts) the contact process generates SARs convex in log-log scale,
with no clearly detectable power-law regime.

Conversely, allowing for variations in the \textit{local population
  size} --as occurs in the stepping stone model-- leads to a monotonic
decrease of the exponent $z$ as the number of individuals per site $M$
is increased. For large values of $M$ we numerically found a reduction
in the exponent $z$ up to a factor $1.4$ with respect to the $M=1$,
voter model, value. Our analytical estimate (\ref{eq:analytics})
suggests that this factor can be actually larger (up to a factor $2$)
when the community size $M$ becomes very large. This is a quite
remarkable result in view of the fact that microorganisms, for which a
description in terms of very large local communities is appropriate
\cite{Fenchel2004-2}, do actually spatially structure themselves with
shallow taxa-area laws \cite{Green2004,horner2004} characterized by
smaller values of $z$.  For instance, a recent review \cite{Microbial}
reports results for salt-marsh bacteria, marine diatoms, arid soil
fungi, and marine ciliates and shows that, in contiguous habitats,
$z$-values for all these categories are roughly the half as those for
larger animals and plants, in surprisingly good agreement with our
results.  It is worthwhile to remark that the reduced diversification
of microorganism has been sometimes ascribed to the possibility of
long distance dispersal \cite{Fenchel2004-2}.  However, numerical
results of Ref.\cite{Rosindell2009} show that distant dispersal events
increase rather than decrease the local slope in the intermediate
regime and thus the value of the exponent $z$ (while the local slope
at larger scales decreases). Therefore, it is unlikely that large
distance dispersal events can --in the absence of additional
mechanisms-- account for the observed small value of the exponents in
microbial communities.

More generally, the spatial variation of biodiversity observed in the
stepping stone model suggests an explanation for the observed
``cosmopolitan'' behavior of microorganisms
\cite{Fenchel2003,Fenchel2004-1,Fenchel2004-2}, where relatively small
areas are found to contain significant fractions of the species known
in the entire globe.  This phenomenon is remarkably well captured by
the SSM where, upon increasing the local population size, each site
tends to contain a considerable fraction of the entire biodiversity
found in a large area.  In conclusion, the results obtained with the
stepping stone model add mathematical support, within the neutral
theory framework, to large population sizes being one of the
mechanisms for the shallower SAR curves observed in microorganisms (as
put forward by Fenchel and Finally
\cite{Fenchel2003,Fenchel2004-1,Fenchel2004-2}). Specifically, the SSM
shows that having a large population size, within a well mixed patch,
provides a buffer to local extinctions and enhances the local
fixations time, making inter-patch migration more effective and the
whole ecosystem closer to a panmictic population with, consequently, a
lowered spatial diversification.

\section*{Supporting Information}
% Title must be 150 characters or less
\renewcommand{\theequation}{S1-\arabic{equation}}
\renewcommand{\thefigure}{S1-\arabic{figure}}
\setcounter{equation}{0}
\setcounter{figure}{0}

\begin{center}{\Large \textbf{Appendix S1}}\end{center}

\begin{flushleft}
\textbf{\Large Dual representation of  voter and stepping stone models}
\end{flushleft}
To compute SAR-curves, one needs to generate several spatial
configurations of the different models and average over
them. Generating such configurations amounts to identify, at a certain
(long) time $t$, to which species the individuals residing at each
lattice site belong to. A straightforward algorithm is to evolve the
system from time $0$ to time $t$ according to the dynamical rules
described in the main text. For the multispecies voter model (MVM) and
stepping stone model (SSM) an alternative strategy exists, that is to
employ the so-called dual dynamics.  This idea stems from the work of
Liggett \cite{Liggett1975}, who recognized the relation between the voter
model without speciation ($\nu=0$) and a system of coalescing random
walkers moving backward in time, and from the coalescent theory
introduced by Kingman \cite{Kingman1982b,Kingman1982a} in the context of
population genetics.

Let us start by briefly recalling the dual representation for the
voter model.  In the voter model an individual creates a replica of
itself at a randomly chosen nearest neighbor site.  In the backward
picture, the sequence of \emph{ancestors} of any given individual,
existing at time $t$, is seen as a random walk moving backward in time
on the lattice. If, at a certain time $s<t$, two random walkers meet
at a site, the corresponding two individuals have a common ancestor
and thus belong to the same species. Hence, for times, $t'<s$ the two
walkers \emph{coalesce} into one. As a consequence, as time evolves
backward, the number of walkers is progressively reduced. Introducing
a non-vanishing speciation rate, $\nu$, corresponds to annihilating
random walkers at the same rate, i.e. to terminating backward
paths. In this way, the voter model with speciation or multispecies voter
model (MVM) turns out to be \emph{dual} to a system of diffusing and
annihilating random walkers moving backward in time.  The dual
representation allows for understanding some properties of the MVM and
SSM in terms of diffusive processes \cite{Durrett-Levin}, and for
deriving analytical predictions (see Appendix S3). Moreover, it suggests
efficient algorithms for numerical simulations. For the MVM, details
of such algorithms can be found in
\cite{Chave2002,Rosindell2007,Rosindell2008,Pigolotti2009}, while
details of our own results for the dual theory of the SSM, together
with a description of an efficient algorithm for computer simulations,
are described in the next subsection.

\renewcommand{\theequation}{S2-\arabic{equation}}
\renewcommand{\thefigure}{S2-\arabic{figure}}
\setcounter{equation}{0}
\setcounter{figure}{0}

\begin{center}{\Large \textbf{Appendix S2}}\end{center}

\begin{flushleft}
\textbf{\Large Dual algorithm for the Stepping Stone Model}
\end{flushleft}
The dual process of the SSM is constructed by following the same idea
as that for the MVM (Appendix S1). The main difference is that in the
SSM each site can host $M$ individuals, and reproduction events can
happen either at the same site with probability $1-\mu$ or at a
neighbor site with probability $\mu$.  In the backward picture, each
site contains $M$ compartments, hosting --at maximum-- one walker
each.  The dynamics starts by placing a walker at each compartment of
a $L\times L$ square-lattice, resulting in $N_w=M\times L^2$
walkers. As for the MVM, the walkers move backward in time coalescing
and annihilating. The only difference here is that coalescence occur
only if the walkers end up in the same compartment within the same
site.

More specifically, the dual algorithm is implemented as follows.  At
each time step, a walker is randomly picked, and: \newline
\textbf{(i)} killed with probability $\nu$.  Then, the number of
species $S$ (set to zero at the beginning) is incremented by one unit
and assigned to the dead walker, which is removed from the pool of
alive walkers ($N_w$ decreased by one).  \newline \textbf{(ii)} with
probability $1-\nu$, the walker moves: \newline \indent\textbf{(iia)}
with probability $1-\mu$, it moves to a randomly chosen compartment
among the other $M-1$ belonging to the same site. If it was occupied,
coalescence takes place, and one of the two coalesced walkers is
removed from the list of alive walkers ($N_w$ decreased by one).
\newline \indent\textbf{(iib)} with probability $\mu$, it moves to a
randomly chosen compartment of any of the neighbor sites. If such
compartment was occupied, coalescence occurs as in \textbf{(iia}).

Each simulation ends when a single walker remains alive ($N_w=1$);
this is then killed as in \textbf{(i)}. Then, the stored information
about the coalescing and annihilating events for each walker allows
the genealogy of each individual to be reconstructed and thus we can
assign a species to each individual, compartments by compartments and
site by site.

The MVM corresponds to $M\!=\!\mu\!=\!1$.  With $M\!=\!1$ and $\mu<1$,
events \textbf{(iia)} correspond to time steps in which walkers do not
move, leading to a time rescaling with respect to the $\mu=1$
case. This clarifies why the effective speciation-to-diffusion ratio,
$\nu/\mu$, is the appropriate parameter to compare the SSM with the
MVM.

Simulations based on the backward dynamics present a number of
advantages.  As both annihilation and coalescence events decrease the
number of walkers, the computation speeds up also when $\nu$ is very
small (where the bottleneck becomes the speed of the random number
generator).  When there are many alive walkers (i.e. at the
beginning), the slowest operation is to search for the
collision/coalescence partner \textbf{(ii)}; this search is made
efficient by means of a look-up table.  Moreover, simulations are free
of finite-size boundary effects.  As walkers can move in the whole
plane, they effectively sample a portion --of size $L\times L$-- of an
infinite system.  This means that, to explore the power-law
intermediate regime of SAR-curves, there is no need to consider huge
systems to avoid finite-size effects, at variance with the forward
dynamics (as discussed in Appendix S4 for the MCP). Finally, there is
no need to wait for a statistically steady state to establish, as is
the case for the forward dynamics.  In the dual representation, each
simulation generates, by construction, a statistically stationary
configuration.

As annihilating and coalescing events are independent, one could in
principle perform simulations at $\nu=0$, and perform killing
(speciation) a posteriori by pruning the list of coalescences. In this
way one could use the same realization of the coalescing random walks
to increase the statistics by averaging over different speciation
histories. This is a standard procedure in population genetics
\cite{Gillespie}, and can also be implemented in this context
\cite{Rosindell2008}.  Here we did not implemented such a procedure
owing to its memory requirements for large $M$ values and large
lattice sizes.

\renewcommand{\theequation}{S3-\arabic{equation}}
\renewcommand{\thefigure}{S3-\arabic{figure}}
\setcounter{equation}{0}
\setcounter{figure}{0}

\begin{center}{\Large \textbf{Appendix S3}}\end{center}

\begin{flushleft}
\textbf{\Large Estimates for  $z$ in the MVM and the SSM}
\end{flushleft}
The backward dynamics can also be used to obtain analytical estimates
for $z$. Before deriving the new results for the SSM, we sketch the
original idea developed for the MVM in
Refs.\cite{Durrett-Levin,Bramson1}.

The timescale for a new species to appear, i.e.  the typical time for
walkers to annihilate in the dual representation, is $\tau=1/\nu$.
Walkers diffuse in space and thus after a time $t$, on average, they
move a distance $\propto t^{1/2}$ from their origin.  This means that,
associated with $\tau$, there is a characteristic spatial scale $\xi =
\sqrt{\tau}=\nu^{-1/2}$. Given a sample of area $\xi^2=\nu^{-1}$, the
number of species present in the sample $S(\xi^2)$ is given by the
total number of annihilated walkers, which can be estimated as
follows.  In a two-dimensional system of coalescing walkers, with
short range dispersal and without annihilation, the density of walkers
decreases asymptotically as \cite{Bramsongriff}
\begin{equation}
\rho(t) \approx \frac{\ln t}{\pi t}\,.
\label{eq:decay1}
\end{equation}
The annihilation rate at time $t$ can be estimated as the annihilation
rate per walker, $\nu$, times the average number of walkers at time
$t$, i.e. $\xi^2 \rho(t)$.  Its time integral gives the total number
of annihilations (the distribution of annihilations is Poissonian), so that
\begin{equation}\label{walkerarg}
S(\xi^2) \sim \nu \xi^2\int_{t_0}^{\tau=\xi^2} \!\!\!\!\!\!\!\!\!\!\! dt\, \rho(t) \sim \frac{\ln^2(\xi^2)-\ln^2(t_0)}{2\pi}
\end{equation}
where $t_0$ is the time at which the asymptotic scaling relation sets
in. We assume $\tau=\xi^2\gg t_0$ and drop the dependence on $t_0$ in
the above expression.  Moreover, the number of annihilations happening
for times larger than $\tau$ is bounded and can be neglected
\cite{Bramson1}.  Finally, the estimate of Eq.~(\ref{walkerarg})
assumes the number of annihilations being negligible compared to the
number of coalescences, i.e.  $\nu$ is very small.

Postulating the scaling form $S(A)\approx N(A)\sim A^{z}$ for
$A\in[1:\xi^2]$ and using the fact that in an area $A=1$ there is
only one species, one obtains \cite{Durrett-Levin}
\begin{eqnarray}\label{dlest}
z = \frac{\ln S(\xi^2)-\ln S(1) }{\ln \xi^2}\sim 
\frac{2 \ln(\ln(\nu^{-1}))-\ln{2\pi}}{ \ln(\nu^{-1})} \approx
\frac{ 2\ln\ln(\nu^{-1}))}{ \ln(\nu^{-1})} \,.\label{z1}
\end{eqnarray}
 We recall that this estimate captures the observed logarithmic dependence of
 $z$ on $\nu$, but it is unable to match the proportionality constant
 computed in numerical simulations \cite{Pigolotti2009}.

Let us now discuss the SSM.  In this case, two different regimes $M
\mu \leq 1$ and $M \mu >> 1$ should be distinguished.  In the former,
all walkers at any given site typically coalesce intra-site before
having the time to jump to neighboring sites, so that essentially no
inter-site coalescences occur before all walkers at any site coalesce
into just one.  Once this has happened, the system becomes voter-like,
and one can repeat the calculation above, but with the diffusion time
replaced by an effective one being $t \mu$. Consequently, one
retrieves the result (\ref{z1}) with $\nu$ replaced by the speciation
to migration ratio: $\bar{\nu}=\nu/\mu$.

In the opposite limit $M \mu >> 1$, intra-site coalescence is limited
by diffusion and by the size of the local population $M$ (after the
initial stage, when the density of walkers has decreased, many
compartments at the same site will be empty so that when two walkers
land on the same site, the probability for them to coalesce is very
small). Since walkers wander for long times before coalescing, we make
the simplifying assumption that all couples of walkers within an area
$\xi^2\approx \bar{\nu}^{-1}$ have the same chance to coalesce,
regardless of their initial separation. In other terms, we assume that
the population living patches of size smaller than $\xi^2$ is well
mixed.  Consequently, we use the mean field formula \cite{MF}
according to which, in the absence of annihilation, the density of
walkers decay as
\begin{equation}
\rho(t) \sim t^{-1}
\label{eq:decay2}
\end{equation}
as opposite to Eq.~(\ref{eq:decay1}). In this case, the equivalent of 
Eq.~(\ref{walkerarg}) becomes
\begin{equation}
S(\xi^2) \sim  \nu M \xi^2 \int_{t_0}^{\tau=\xi^2}
\!\!\!\!\!\!\!\!\!\!\!\! dt\, t^{-1} \sim M\mu
[\ln(\xi^2)-\ln(t_0)]\,.
\label{N2}
\end{equation}
\begin{figure}[!ht]
\begin{center}
\includegraphics[width=0.7\textwidth]{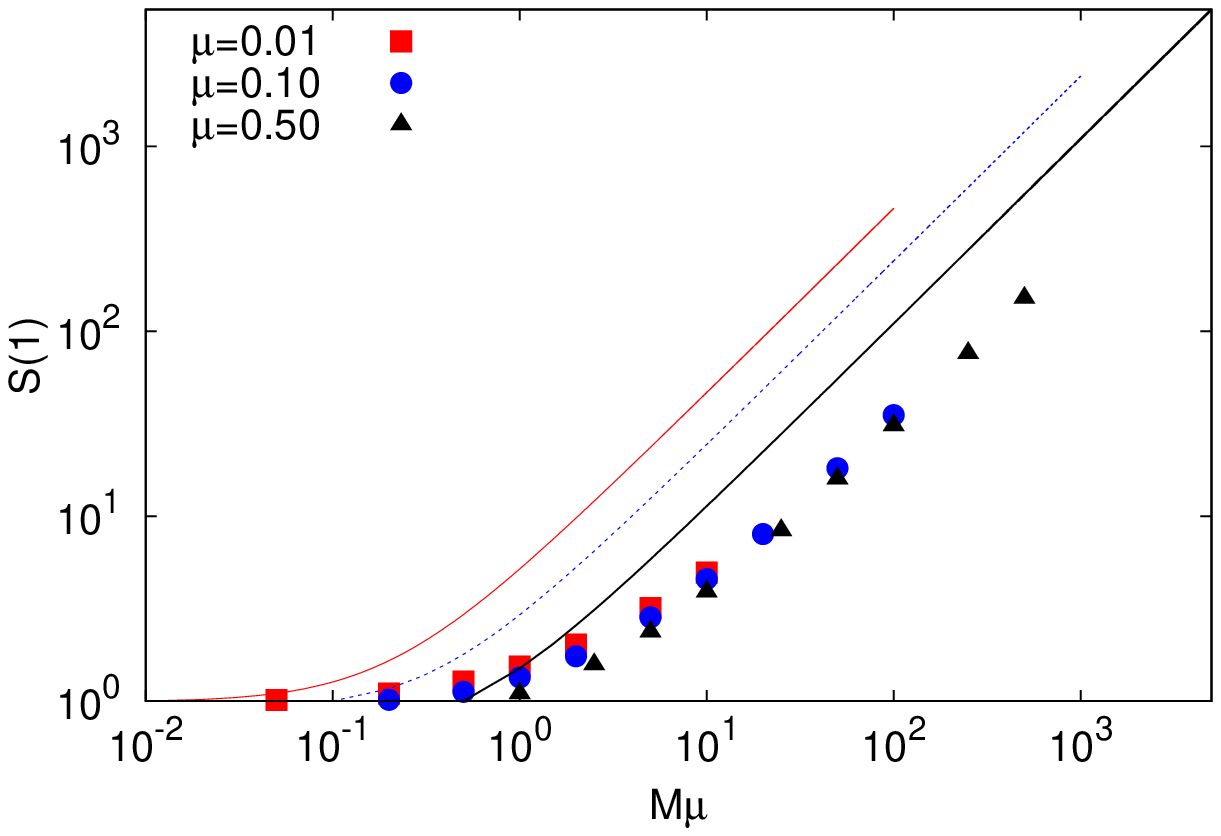}
\end{center}
\caption{ \textbf{Average number of species in a single site of the
    SSM} with $\bar{\nu}=\nu/\mu=10^{-8}$ (parameters as in Fig.~4 of
  the main text) as a function of $M\mu$, for $\mu=0.01, 0.1$ and
  $0.5$ (as labeled). The solid lines are obtained from
  Eq.~(\ref{eq:ewens}), where the colors correspond to the values of
  $\mu$ chosen in the simulations. Numerical simulation and theory
  display a linear behavior for large $M\mu$. The quantitative
  agreement between the prefactors decreases as $\mu$ is decreased as
  low values of $\mu$ generate correlations in the sample not captured
  by formula (\ref{eq:ewens}).  Finally, notice that simulations
  suggest $S(1)$ being a function of $M\mu$ only while in the sampling
  formula $S(1)$ at fixed $M\mu$ still shows a dependence on $\mu$,
  see text for a discussion.  }
\label{fig:ewens}
\end{figure}

In order to compute $z$, we  also need an estimate for $S(1)$, which in
this case is not fixed to be $1$ like in the voter model. As we are
assuming the population in an area $\xi^2$ to be well mixed, we can
think of a single site as a sample of $M$ individuals from this
population and make use of Ewens' sampling formula
\cite{Ewens,Tavare,DurrettBook}. As derived, e.g., in \cite{Tavare}
we have that
\begin{equation}
S(1)=\sum_{j=0}^{M-1} \frac{\theta}{\theta+j}\approx M\mu\log(1+\mu^{-1})
\label{eq:ewens}
\end{equation}
where $\theta$ is the product of the panmictic population size
($M\xi^2$) times the speciation rate $\nu$, i.e. $\theta=
M\xi^2\nu=M\mu$, and the last expression has been derived by
approximating the sum by an integral. Proceeding as in
Eq.~(\ref{dlest}), using Eq.~(\ref{N2}) and Eq.~(\ref{eq:ewens}), we
find
\begin{equation}
z \sim \frac{\ln(\ln(\bar{\nu}^{-1}))-\ln{C}}{ \ln(\bar{\nu}^{-1})}\approx 
\frac{ \ln(\ln(\bar{\nu}^{-1}))}{ \ln(\bar{\nu}^{-1})}, 
\label{z2}
\end{equation}
where $C\approx \log(1+\mu^{-1})$. Notice how, in the limit of small
$\bar{\nu}$, $z$ is a factor $2$ smaller than the prediction of
Eq.~(\ref{z1}) for the voter model.

We conclude by remarking that the panmictic behavior of the SSM is a
rigorous result when periodic boundary conditions are implemented on a
square of side $L$ and $ M \mu \gg\ln(L)$, as shown in
\cite{CoxDurrett2002}. Conversely, in the open boundary condition
case considered here it must be taken as an approximation, whose
accuracy may depend on the spatial scale. We tested this approximation
by comparing the estimate of the average number of species in one site
from numerical simulation with the prediction based on the Ewens'
sampling formula. Results are presented in Fig.~\ref{fig:ewens},
showing that the linear behavior in $M\alpha$ is well predicted by
formula (\ref{eq:ewens}), apart from a difference in the prefactor. In
particular, simulations suggest a prefactor $C\approx 0.3$, which
seems to be independent of $\mu$ (so that $S(1)$ becomes a function of
$M\mu$ only), while the estimate of Eq.~(\ref{eq:ewens}) predicts a
constant dependent on $\mu$ which deviates more from the numerical
results as $\mu$ is decreased. The reason of this deviation can be
ascribed to the effect of inter-site coalescence at smaller values of
$\mu$, reducing the number of species compared to the theory,
i.e. compensating the increase of $M$ (at $M\mu$ fixed). Another way
of seeing it is that individuals belonging to the same site at low
$\mu$ constitute a correlated sample of the population.  Conversely,
in Ewens prediction (\ref{eq:ewens}) when $M$ increases (at fixed
$M\mu=\theta$) it means that the sample size increases leading,
obviously, to a larger number of species.

The issue above demonstrates a problem common also to the estimates
for the MVM: quantitative agreement between theory and simulations can
be compromised by hard-to-estimate constants, whose contribution
becomes irrelevant only for inaccessibly small values of $\nu$ (where
the only relevant contribution is the $\ln \ln \nu$ term).

\renewcommand{\theequation}{S4-\arabic{equation}}
\renewcommand{\thefigure}{S4-\arabic{figure}}
\setcounter{equation}{0}
\setcounter{figure}{0}

\begin{center}{\Large \textbf{Appendix S4}}\end{center}

\begin{flushleft}
\textbf{\Large Numerical implementation of the MCP}
\end{flushleft}
The contact process is a self-dual model \cite{Liggett1985}, meaning
that its dual representation is the contact process itself. Therefore,
duality does not provide any useful help in this case, and one needs
to resort to standard forward-time simulations. Here below we briefly
sketch the algorithm we used and discuss some related issues.

We implemented a multispecies generalization of standard algorithms
for the contact process \cite{MarroBook}.  A $L\times L$
two-dimensional lattice with periodic boundary conditions is
initialized by placing and individual at each lattice site; its
species is labeled by a positive integer $s$ ($s=0$ means that the
site is empty).  Initially, a single species occupies the whole
system.  We keep track of occupied sites in a list, containing
$N_{occ}=L^2$ entries at time $t=0$. At each step, time is incremented
by $L^2/N_{occ}$ and a random individual (in the list of non-empty
sites) is chosen:\newline \textbf{(i)} with probability
$\delta/(\beta\!+\!\delta)$ it is killed and removed from the list
($N_{occ}$ decreases in one unit) \newline \textbf{(ii)} with
probability $\beta/(\beta+\delta)$, reproduction at a randomly chosen
neighbor site is attempted: \newline \indent\textbf{(iia)} if the
chosen neighbor site was non-empty, reproduction is unsuccessful, and
the state of the system does not change \newline \indent\textbf{(iib)}
if it was empty, reproduction is successful ($N_{occ}$ increases in
one unit) \newline \indent\hspace{.1cm} \textbf{(iib.1)} with
probability $1-\nu$ the site becomes occupied by an individual from
the parent species and is added to the list of occupied sites;
\newline \indent\hspace{.1cm}\textbf{(iib.2)} with probability $\nu$
the newborn mutates from the parent, giving birth to a new species. A
new species-label is created by increasing in unity the largest
existing one, and it is assigned to this site (which, on its turn, is
added to the list of non-empty sites).

When $\gamma=\beta/\delta>\gamma_c$, the system evolves to a
dynamical equilibrium, with new species appearing and older ones
becoming eventually extinct.  The number of extant species is
monitored. Once equilibration of the density of non-empty sites is
reached, on longer time scales (on the order of $\nu^{-1}$) also the
number of extant species equilibrate fluctuating around a mean value.
After equilibration, configurations of the system are periodically
sampled, and used to compute the SAR. Notice that, in order to
have statistically independent measurements, the sampling interval
should be also on the order of $\nu^{-1}$. Simulations are terminated
when enough statistics have been collected.

At variance with the backward algorithm for the SSM and MVM, here
boundaries play and important role, requiring rather large lattices to
avoid finite size effects. In particular, $L$ must be larger than
$1/\sqrt{\nu}$, which sets (similarly to the SSM and MVM) an
approximate scale on which individuals are expected to diffuse before
speciation. If $L$ is taken too small, the measured effective value
$z$ underestimates the true one.  By comparing SARs obtained at equal
parameters and different system sizes, we determined that a safe
choice to neglect finite size effect is $L^2=10/\nu$. We also checked
that for this system size and large values of $\gamma$ the results for
the MVM are recovered.

Consequently, at decreasing $\nu$ simulations become more and more
demanding both because a larger system size is required and because
relevant time scales become slower and slower. Owing to these
limitations, we could not simulate systems with $\nu$ smaller than
$10^{-6}$ with enough statistics.

\renewcommand{\theequation}{S5-\arabic{equation}}
\renewcommand{\thefigure}{S5-\arabic{figure}}
\setcounter{equation}{0}
\setcounter{figure}{0}

\begin{center}{\Large \textbf{Appendix S5}}\end{center}

\begin{flushleft}
\textbf{\Large Behavior of bare
  average for the MCP at low densities and low speciation rates}
\end{flushleft}
We discuss in detail how approaching the critical birth-to-death rate
ratio, $\gamma\to \gamma_c$, SARs measured using the bare average
(Eq.~(4) of the main text) are mostly determined by spatial
fluctuations of the density of individuals rather than by species
diversity.

For $\gamma=\gamma_c$, the set of occupied sites constitutes a fractal
set --- with fractal-dimension $d_F<d$ --- embedded in the
$d$-dimensional space (see, e.g. \cite{MarroBook}). In this limit, the
coarse-grained density $\rho(A)$ of regions of area $A$ occupied by at
least one individual grows as $\rho(A)\sim A^{d-d_F}$.  As a
consequence, when $\nu$ is very small and bare averages are chosen,
one has $S(A)\approx \rho(A)\sim A^{d-d_F}$ for small areas, leading
to an estimate $z=d-d_F$.

\begin{figure}[!ht]
\begin{center}
\includegraphics[width=0.7\textwidth]{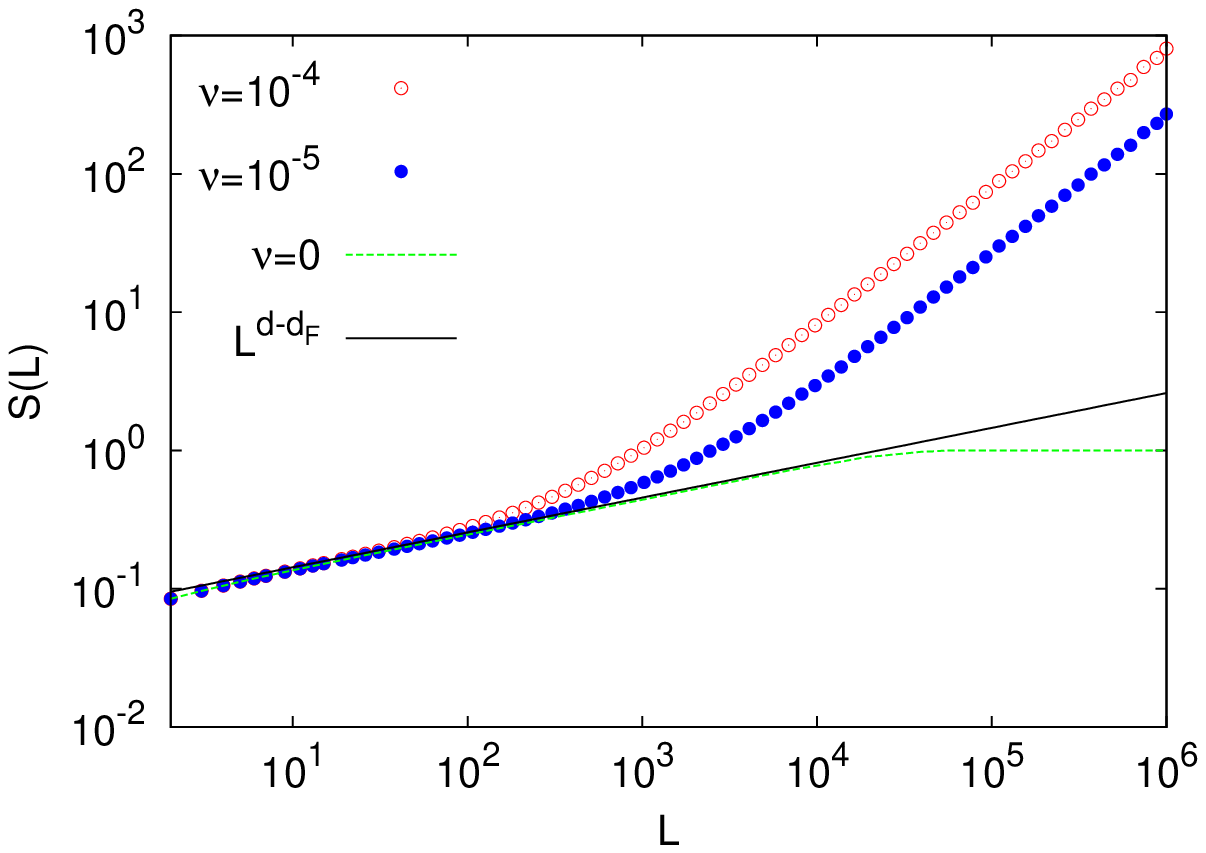}
\end{center}
\caption{ \textbf{Species-length relationship for the 1D contact process
    close to the critical birth-to-death rate ratio},
  $\gamma_c-\gamma=10^{-4}$ (in $d=1$ $\gamma_c=3.29785(2)$) for
  $\nu=10^{-4}$ (red symbols) and $10^{-5}$ (blue symbols), in a
  lattice of size $10^6$. Bare averages (Eq.~(4) of the main text) are
  employed to compute the number of species $S(L)$ over segments of
  length $L$. At $\gamma_c-\gamma=10^{-4}$ criticality becomes
  evident: occupied sites live on a fractal set of dimension
  $d_F\approx 0.75$, which implies a spurious power-law behavior
  $S(L)\sim L^z$ with $z=d-d_F\approx 0.25$, as shown by the black
  straight line. Also shown for comparison is the bare average with
  $\nu=0$ (only one species). Notice how this coincides with the
  $\nu\neq 0$ SLR for $L$ short enough, confirming that the power-law
  has a pure geometrical origin. }
\label{fig:cp1d}
\end{figure}

Unfortunately, in two dimensions $d=2$, it is very hard to verify this
prediction, due to the difficulties in simulating MCP close to
$\gamma_c$ and for small values of $\nu$. Here, we demonstrate this
effect in the numerically simple one-dimensional case.  In $d=1$, one
has $d_F\approx 0.75$ (see, e.g. \cite{MarroBook}), and the
previous argument predicts $z=d-d_F\approx 0.25$. This is confirmed in
Fig.~\ref{fig:cp1d} where we show $S(A)$ for $\nu=0,10^{4}$ and
$10^{-5}$ for $\gamma_c-\gamma=10^{-4}$.
%%%FIg2 SUPP WAS HERE

Finally, we remark once more that in the framework of SARs this regime
must be considered as an artifact induced by bare averages, in the
sense that the resulting power law does not contain information about
species diversity. This is made more clear in the figure, where the
same power law is observed for $\nu=0$ where only one species is
present.

% Do NOT remove this, even if you are not including acknowledgments
\section*{Acknowledgments}
We thank J. Rosindell for useful comments and suggestions.  M.A.M.
thanks the Spanish MICINN project FIS2009-08451 and Junta de
Andaluc{\'\i}a, Pr. Excelencia P09-FQM4682 for financial support.  We
also thank the GENIL program of the University of Granada for support.
The funders had no role in study design, data collection and analysis,
decision to publish, or preparation of the manuscript.

%\section*{References}
% The bibtex filename
%\bibliography{bibtot}

\end{document}